\documentclass[aps,showpacs,floatfix,superscriptaddress,twocolumn]{revtex4-1}
\usepackage{enumitem}
\usepackage{graphicx}
\usepackage{amssymb}
\usepackage{float}
\usepackage{graphics,color}
\usepackage{amsmath}
\usepackage{subfigure}
\usepackage{amsfonts}
\usepackage{hyperref}
\usepackage{epstopdf}
\usepackage{graphicx,psfrag}
\DeclareGraphicsExtensions{.pdf,.eps,.png,.jpg,.mps} 

\def\gsim{\, \rlap{$>$}{\lower 1.1ex\hbox{$\sim$}}\,}
\def\lsim{\, \rlap{$<$}{\lower 1.1ex\hbox{$\sim$}}\,}

\makeatletter

\newcommand\numberthis{\addtocounter{equation}{1}\tag{\theequation}}

\begin{document}
\title{Number theory, periodic orbits and mean-field superconductivity in nano-cubes}
\author{James Mayoh}
\affiliation{University of Cambridge, Cavendish Laboratory, JJ Thomson Ave., Cambridge, CB3 0HE, UK}
\author{Antonio M. Garc\'{\i}a-Garc\'{\i}a}
\affiliation{University of Cambridge, Cavendish Laboratory, JJ Thomson Ave., Cambridge, CB3 0HE, UK}
\affiliation{CFIF, Instituto Superior T{\'e}cnico,
Universidade de Lisboa, Av. Rovisco Pais, 1049-001 Lisboa, Portugal}
\begin{abstract}
We study superconductivity in isolated superconducting nano-cubes and nano-squares of size $L$ in the limit of negligible disorder,  $\delta/\Delta_0 \ll 1$ and  $k_F L \gg 1$ for which mean-field theory and semiclassical techniques are applicable, with $k_F$ the Fermi wave vector, $\delta$ the mean level spacing and $\Delta_0$ the bulk gap. By using periodic orbit theory and number theory we find explicit analytical expressions for the size dependence of the superconducting order parameter. Our formalism takes into account contributions from both the spectral density and the interaction matrix elements in a basis of one-body eigenstates. The leading size dependence of the energy gap in three dimensions seems to be universal as it agrees with the result for chaotic grains. In the region of parameters corresponding to conventional metallic superconductors, and for sizes $L \gtrsim 10$nm, the contribution to the superconducting gap from the matrix elements is substantial ($\sim 20\%$). Deviations from the bulk limit are still clearly observed even for comparatively large grains $L \sim 50$nm. These analytical results are in excellent agreement with the numerical solution of the mean-field gap equation.
\end{abstract}
\maketitle
The evolution of superconductivity in confined geometries as the grain size enters the nano-scale region has been a recurrent research theme for more than fifty years. Anderson\cite{Anderson1959} was the first to note, in the late fifties, that superconductivity should be strongly disturbed as the mean level spacing becomes comparable with the superconducting energy gap. It was not until the experiments on isolated Al nano-grains\cite{Ralph1995,*Black1996} that it was possible to study superconductivity in single nano-grains with relatively good experimental control. The recent experimental observation\cite{Bose2010,*Brihuega2011} of superconductivity in single isolated Sn and Pb hemispherical nano-grains $L \leq 30$nm has confirmed that deviations from the bulk limit can be important even in the limit of relatively large grains, $L \sim 10$nm, where a mean field approach is applicable.

Theoretically it was soon realised that the critical temperature in superconducting nano-cubes\cite{parmenter1968a}, obtained by solving the Bardeen-Cooper-Schrieffer (BCS)\cite{Bardeen1957} gap equation, could be much higher than in the bulk limit for grains where the Fermi energy was in a region with an anomalously large density of states.

Size effects in BCS mean-field theory not only depend on the spectral density around the Fermi energy but also \cite{Bardeen1957} on the 
interaction matrix elements in the basis of one-body eigenfunctions, $I_{{n},{n'}}=V\int \psi_{\bf{n}}^2({\bf r})\psi_{\bf{n'}}^2({\bf r})\,dV$ with $\Psi_n(r)$ the solution of the Schroedinger equation in the grain. In the context of thin films it was shown in\cite{blatt1963,*Thompson1963} that, on average, this contribution always enhances superconductivity. The leading finite size correction related to these matrix elements for chaotic grains \cite{Farine2003} is comparable to that coming from the spectral density. A complete analytical expression\cite{Garcia-Garcia2008} of the size dependence of the superconducting gap for chaotic grains, including  spectral density and matrix element contributions, was found in Ref. \cite{Garcia-Garcia2008}. The semi-classical techniques\cite{Brack1997} employed in Ref.\cite{Garcia-Garcia2008} have also been used to estimate\cite{Olofsson2008} the typical deviation of the superconducting gap from the bulk limit as a function of the grain size and symmetry. 

Numerical studies of single superconducting nano-grains of different geometries -- spheres\cite{Boyaci2001,*Gladilin2006,*Tempere2005}, cylinders\cite{shanenko2006,*Croitoru2007} and harmonic oscillators\cite{heiselberg2003} -- have confirmed the important role played by both the spectral density and the matrix elements in the evolution of the superconducting gap. For $L \geq 10$nm it has been found that size effects are important but still a mean field approach is accurate since the bulk gap is much larger than the mean level spacing. Moreover in this region of sizes the solutions of the Bogoliubov-deGennes equations and the simpler approach of including the matrix elements in the BCS theory, employed in Ref.\cite{Garcia-Garcia2008,Farine2003,Olofsson2008}, lead to similar results\cite{shanenko2006,*Croitoru2007}.

As was mentioned previously analytical studies that combine the effect of the spectral density and matrix elements within a BCS mean-field approach, are restricted to chaotic grains \cite{Garcia-Garcia2008}. It would be interesting to extend this analysis to highly symmetric cubic and spherical grains where greater deviations from the bulk limit are expected. This paper is a step in this direction.

We solve analytically the BCS gap equation for a cubic and square grain by using periodic orbit theory. Our main result is an expansion, in the semiclassical parameter $(k_FL)^{-1}$, of the superconducting gap, that takes into account corrections due to both the matrix elements and the spectral density. We show that in the region of interest $L \gtrsim 10$nm, for which BCS is still applicable, the matrix element contribution is substantial. For metallic grains of some weakly coupled BCS superconductors noticeable deviations from the bulk limit are still observed for $L \sim 50$nm. We start by introducing the model and the techniques employed in our theoretical analysis. 

\section{The model}
BCS theory describes pairing between electrons by a Hamiltonian of the form\cite{Bardeen1957},
\begin{equation}
H=\sum_{{\bf n}\,\sigma}\epsilon_{\bf n} c^\dag_{{\bf n}\sigma}c_{{\bf n}\sigma}-\frac{\lambda}{\nu(0)}\sum_{{\bf n},{\bf n'}}I_{{\bf n},{\bf n'}}c_{{\bf n}\uparrow}^\dag c_{{\bf n}\downarrow}^\dag c_{{\bf n'}\uparrow}c_{{\bf n'}\downarrow}
\end{equation}
where $c_{{\bf n}\sigma}^\dag$ creates an electron of spin $\sigma$ in a state with quantum numbers ${\bf n}$ and energy $\epsilon_{\bf n}$, $\lambda$ is the dimensionless BCS coupling constant for the material and $\nu(0)$ is the density of states at the Fermi energy. The short range electron-electron interaction matrix elements are given by,
\begin{equation}
I_{{\bf n},{\bf n'}}=V\int \psi_{\bf{n}}^2({\bf r})\psi_{\bf{n'}}^2({\bf r})\,dV
\label{Mel}
\end{equation}
where V is the volume of the grain and  $\psi_{\bf{n}}({\bf r})$ is the eigenfunction of the one-body problem labeled by the quantum numbers $\bf{n}$.\\
The BCS order parameter is defined by,
\begin{equation}
\Delta_{\bf{n}}= \frac{\lambda}{\nu(0)}\sum_{{\bf n'}}I_{{\bf n},{\bf n'}}\langle c_{{\bf n'}\uparrow}^\dag c_{{\bf n'}\downarrow}^\dag \rangle,
\end{equation}
and can be calculated from the self-consistency equation,
\begin{equation}
\Delta_{\bf n}=\frac{\lambda}{2}\displaystyle\sum_{{\bf n'}}\frac{\Delta_{{\bf n'}}I_{{\bf n},{\bf n'}}}{\sqrt{\epsilon_{{\bf n'}}^2+\Delta_{{\bf n'}}^2}}\frac{1}{\nu(0)}
\label{GapEqn}
\end{equation}
where the sum is now taken over all elements of the set $\left\{{\bf n'}\big|\:|\epsilon_{{\bf n'}}|<\epsilon_D \right\}$, where $\epsilon_D$ is the Debye energy\cite{footGE}.
In the bulk limit and for negligible disorder the eigenfunctions are well approximated by simple plane waves  so that $I_{{\bf n},{\bf n'}} \approx 1$ which leads to the well known relation for the superconducting gap,
\begin{equation}
\Delta_0 \approx 2\epsilon_De^{-\frac{1}{\lambda}}.
\end{equation}
 However in small grains one expects $I_{n,n'}$ can deviate significantly from its bulk value. Here we consider the enhancement of the gap due to the matrix elements in small grains.
We restrict our interest to grains in which both a mean-field BCS theory, $\delta/\Delta_0 \ll 1$, with $\delta = 1/\nu(0)$ the mean level spacing at the Fermi energy, and the semi-classical periodic orbit theory, $k_FL\gg1$, are applicable.

For our system of interest, a cubic or square grain, the eigenfunction of the one-body problem are simply,
\begin{equation}
\psi_{\bf n}({\bf r})=
\begin{cases}
\frac{2}{\sqrt{A}}\sin(\frac{n_x\pi}{L_x}x)\sin(\frac{n_y\pi}{L_y}y), &\text{(2D)}\\
\frac{2\sqrt{2}}{\sqrt{V}}\sin(\frac{n_x\pi}{L_x}x)\sin(\frac{n_y\pi}{L_y}y)\sin(\frac{n_z\pi}{L_z}z),&\text{(3D)}
\end{cases}
\label{RecWF}
\end{equation}
with eigenenergies
\begin{equation}
\epsilon_{\bf n}=
\begin{cases}
\frac{\hbar^2 \pi^2}{2m}\left(\left(\frac{n_x}{L_x}\right)^2+\left(\frac{n_y}{L_y}\right)^2\right), &\text{(2D)}\\
\frac{\hbar^2 \pi^2}{2m}\left(\left(\frac{n_x}{L_x}\right)^2+\left(\frac{n_y}{L_y}\right)^2+\left(\frac{n_z}{L_z}\right)^2\right),&\text{(3D).}
\end{cases}
\end{equation}
where, $L_x,L_y,L_z$ are the side lengths of the grain, A(V) is the area(volume) and ${\bf n}= (n_x,n_y,n_z)$ are not simultaneously zero. The matrix element can be easily calculated,
 \begin{equation}
I_{\bf n,n'}=
\begin{cases}
(1+\frac{1}{2}\delta_{n_xn_x'})(1+\frac{1}{2}\delta_{n_yn_y'}), &\text{(2D)} \\
(1+\frac{1}{2}\delta_{n_xn_x'})(1+\frac{1}{2}\delta_{n_yn_y'})(1+\frac{1}{2}\delta_{n_zn_z'}), &\text{(3D)}
\end{cases}
\label{matrixEls}
\end{equation}
where $\delta_{\alpha,\beta}$ is the Kronecker delta. The problem of computing the matrix element is equivalent to that of finding the number of shared quantum numbers for a given state, which is closely related to the of the degeneracy of a given energy level. The latter are usually referred to as shell effects. Here we will consider the special case where $L_x=L_y=L_z=L$ in which the level degeneracy is higher and hence we expect stronger size effects. For clarity we will use the word `state' exclusively to refer to a single electron state of the system and the word `shell' to refer to the full set of degenerate states at some energy. 

In next section \ref{sec2} we find an analytical expression for the average size dependence of the superconducting gap by applying results from number theory, more specifically we relate the behavior of the Diophantine equation, $n_x^2+n_y^2+n_z^2=n$ to the problem of level degeneracies in a cubic grain. In section \ref{sec3} we employ periodic orbit theory, valid in the semiclassical limit $k_FL \gg 1$, in order to find an analytical expression for the non-monotonic size dependence of the superconducting gap. These analytical expressions include finite-size contributions from both the matrix elements and the spectral density.
\begin{figure}[t]
    \includegraphics[width=0.45\textwidth]{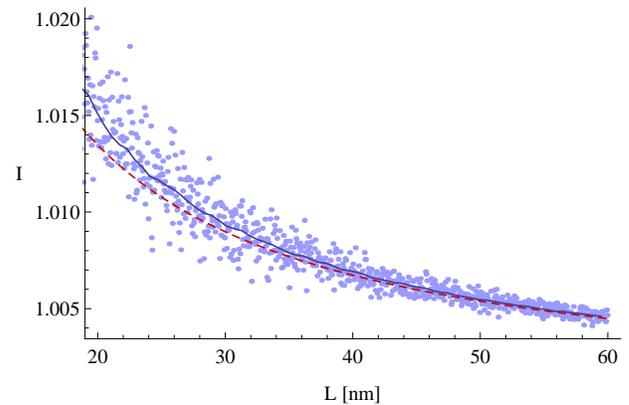}
 \caption{The matrix elements Eq.(\ref{matrixEls}) as a function of the square size $L$. Blue dots correspond to the exact numerical calculation, the blue solid line shows the numerical average of the exact results and the red dashed line corresponds to the analytic prediction, Eq.(\ref{ME3Dsmooth}). For sufficiently large grains $L >20$nm the agreement is excellent. For small sizes deviations are expected since our results neglect higher order terms in the expansion parameter $(k_FL)^{-1}$.}\label{3dME}
\end{figure}
\section{Average size dependence of the gap: results from Number Theory}\label{sec2}
In this section we apply results from number theory to study the mean size dependence of the matrix elements and the superconducting gap. To make the problem tractable we will remove the $n$ dependence from the left hand side of Eq.(\ref{GapEqn}) by replacing $I_{n,n'}$ with its average taken across all possible states in the Debye window and Fermi level for $n$ and $n'$ respectively. The gap equation can then be solved by taking the matrix element outside the integral.

The gap equation may be written in the more transparent form,
\begin{equation}\label{Gap2}
\Delta_{\bf n}=\frac{\lambda}{2}\displaystyle\sum_{ n'}\frac{\Delta_{n'}I_{{\bf n}, n'}}{\sqrt{\epsilon_{ n'}^2+\Delta_{ n'}^2}}\frac{1}{\nu(0)}
\end{equation}
where the sum is over the set $\{n'; |\epsilon_{n'}-\epsilon_f|<\epsilon_D\}$ and we have moved the sum over quantum numbers into the definition of the matrix element as this is the only term which depends upon them explicitly. Therefore,
\begin{equation}
I_{{\bf n}, n'} \equiv \sum_{\{n'\}} I_{{\bf n},{\bf n'}}
\end{equation}
where this sum is over the quantum numbers $\{(n'_x,n'_y,n'_z);n'={n'_x}^2+{n'_y}^2+{n'_z}^2\}$. This is the starting point for the number theory analysis. 
\begin{figure}[t]
    \includegraphics[width=0.45\textwidth]{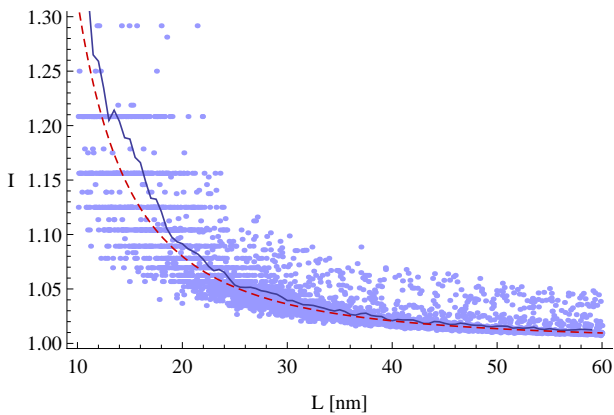}
 \caption{The matrix elements Eq.(\ref{matrixEls}) as a function of the square size $L$. Blue dots correspond to the exact numerical calculation,  the blue solid line shows the numerical average of the exact results and the red dashed line corresponds to the analytic prediction, Eq.(\ref{two_dimensional_ME}). Since the analytical calculation does not neglect any term in the expansion the agreement is much better than for the cube across the whole size range. 
 }\label{2dME}
\end{figure}

\begin{widetext}
\subsection{Three Dimensions}
For a given state with quantum numbers ${\bf n}=(n_x,n_y,n_z)$ we can calculate the total matrix element due to states in the Debye window by expanding Eq.(\ref{matrixEls}), 
\begin{equation}
I_{{\bf n}, n'}=\sum_{\{n'\}}\left(1+\frac{1}{2}\sum_{r\in\{x,y,z\}}\delta_{n_r,n_r'}+\frac{1}{4}\sum_{\langle r,s\rangle}\delta_{n_r,n_r'}\delta_{n_s,n_s'}+\frac{1}{8}\delta_{n_x,n_x'}\delta_{n_y,n_y'}\delta_{n_z,n_z'}\right)
\end{equation}
where $\langle r,s\rangle \equiv (r,s)\in\{(x,y),(y,z),(z,x)\}$. The sum in $\{n'\}$ is then carried out leading to, 
\begin{equation}\label{ME1}
I_{{\bf n}, n'}=r_3(n')+\frac{1}{2}\sum_{r\in\{x,y,z\}}r_2(n'-n_r^2)+\frac{1}{4}\sum_{\langle r,s\rangle}r_1(n'-n_r^2-n_s^2)+\frac{1}{8}
\end{equation}
\end{widetext}
where $r_i(n)$ is the number of non-negative representations of $n$ as the sum of $i$ squares. For example $r_3(n)$ is the number of solutions to the Diophantine equation $n=a^2+b^2+c^2$ such that $a,b,c$ are non-negative integers. In other words it is the degeneracy of the $\epsilon_n$ shell in a cube. In the appendix \ref{ap:num} we provide a summary of the closed forms for these functions using number theory techniques. 
This expression for the matrix elements, together with Eq. (\ref{Gap2}) and appendix \ref{ap:num},  provides a complete description for the system. More insight can be gained by some further re-arrangements. First taking the degeneracy of the shell out of the matrix element by defining the mean matrix element,
\begin{equation}\label{ME3Ddef}
{\bar I}_{{\bf n}, n'} \equiv \frac{I_{{\bf n}, n'}}{r_3(n')}
\end{equation}
which facilitates the writing of the gap equation using the familiar integral notation,
\begin{equation}
\Delta_{\bf n}=\frac{\lambda}{2}\int_{-\epsilon_D}^{\epsilon_D}\frac{\Delta_{n'}\bar{I}_{{\bf n}, n'}}{\sqrt{\epsilon^2+\Delta_{ n'}^2}}\frac{\nu(\epsilon)}{\nu(0)}d\epsilon.
\end{equation}
For a cube this problem has been studied and solved exactly \cite{Garcia-Garcia2011} in the limit $I=1$ by using periodic orbit theory \cite{Brack1997}. In order to tackle the problem of a non-trivial ${\bar I}_{{\bf n}, n'} \neq 1$ we carry out a further smoothing over the Debye window by taking an additional average over $n'$, $r_i(n'+\ldots) \rightarrow\sum_{n'}r_i(n'+\ldots)$. The leading finite size correction is then given by,
\begin{equation}
\bar I=1+\frac{\frac{1}{2}\sum_{n'}\sum_{r\in\{x,y,z\}}r_2(n'-n_r^2)}{\sum_{n'}r_3(n')}.
\end{equation}
The mean value of $r_2(n)$ is $\pi/4$ \cite{Grosswald1985} and it is straightforward to show that $\sum_{n'}r_3(n')=k_D^2k_FL^3/2\pi^2$ and $\sum_{n'}1=2k_D^2L^2/\pi^2$. Combining these results we find,
\begin{equation}\label{ME3Dsmooth}
\bar I=1+\frac{3\pi}{2}\frac{1}{k_FL}.
\end{equation}
The correction found here is of the same order as that coming from the Weyl expansion of the density of states. Therefore both contributions must be considered on equal footing. For instance, for Neumann boundary conditions, the correction from the Weyl expansion cancels exactly the contribution from the matrix elements Eq.(\ref{ME3Dsmooth}). We note that for 3d chaotic grains the leading corrections to the matrix elements \cite{Garcia-Garcia2008} is also given by Eq.(\ref{ME3Dsmooth}). This strongly suggests that it is universal, namely, does not depend on the shape of the grain. Analytical and numerical results for Dirichlet boundary conditions, depicted in figure \ref{3dME}, are in good agreement for $L \geq 20$nm. For smaller sizes deviations are indeed expected since the analytical calculation only provides the leading size correction. Explicit expressions for higher orders are hard to obtain by number theory techniques. 

\subsection{Two Dimensions}
As in three dimensions we write the mean matrix element as,
\begin{equation}
\begin{split}
I_{{\bf n}, n'}&=\sum_{\{n'\}} \left(1+\frac{1}{2}\sum_{r\in\{x,y\}}\delta_{n_rn_r'}+\frac{1}{4}\delta_{n_xn_x'}\delta_{n_yn_y'}\right) \\
&= r_2(n')+\frac{1}{2}\sum_{r\in\{x,y\}}r_1(n'-n_r^2)+\frac{1}{4}.
\end{split}\label{Inn_2d}
\end{equation}
As before Eq. (\ref{Inn_2d}), along with the number theoretic results in appendix \ref{ap:num}, provides an exact description of the matrix elements.
To study the behavior assuming local smoothing we must consider the contribution from states where just one the quantum numbers match, for example $n_x=n_x', n_y \neq n_y'$. This contribution can be determined probabilistically using the following argument. The states which verify this condition and which are contained in the Debye window are those such that,
\begin{equation}
\Delta\epsilon=\left|\frac{\hbar^2\pi^2}{2mL^2}(a^2\pm2n_xa)\right|<\epsilon_D
\end{equation}
where $a$ is an integer greater than zero. The total possible contribution if all $n_x$ were available for a given grain size is then,
\begin{equation}
T\equiv2\sum_{a=1}^{\sqrt{\sigma^2+1}-1}\sum_{n=1}^{\frac{\sigma^2-a^2}{2a}}1\approx\frac{1}{2}(2\gamma-1+\ln(\sigma))\sigma
\end{equation}

where $\sigma=(\frac{k_DL}{\pi})^2$,  the factor two accounts for positive and negative values, and $\gamma\approx0.577$ is the Euler-Mascheroni constant.
However not every $n_x$ is available in the shell at the Fermi level. In order to determine the fraction of available $T$ which is realized we note that, provided that every $n_x<\sqrt{n-1}$ has equal probability to be in the Fermi level shell, there are approximately $\frac{k_FL}{\pi}$ possible values for $n_x$ of which, discounting permutations, $r_2(n)/2$ independent values are chosen. Hence the total contribution to the shell from states of this form is,
\begin{equation}
T\frac{\pi r_2(n)}{2k_FL}.
 \end{equation}
 As a result,
\begin{equation}\label{2d_num_res}
\bar I=1+\frac{1}{\sum_{n'}r_2(n')}\left(\frac{5}{4}+T\frac{\pi^2}{8k_FL}\right)
\end{equation}

where the first term accounts for the case $n_x=n'_x,n_y=n_y'$. The denominator $\sum_{n'}r_2(n)$ is simply the number of single electron states in the Debye window. We calculate this by dividing the phase space volume of the Debye window by the phase space volume of a single electron state. In the $\epsilon_D \ll \epsilon_F$ limit,
\begin{equation}
\label{Nt2}
\sum_{n'}r_2(n')=\frac{(k_DL)^2}{2\pi}
\end{equation}
which leads to our final result,
\begin{equation}\label{two_dimensional_ME}
\bar I=1+\frac{5\pi}{2}\frac{1}{(k_DL)^2}+\left(2\gamma-1+ 2\ln\left(\frac{k_DL}{\pi}\right)\right)\frac{\pi}{4k_FL}.
\end{equation}
 The presence of the Euler-Mascheroni constant indicates the inherently number theoretic nature of this result. 

We have found, see figure \ref{2dME}, and excellent agreement between Eq.(\ref{two_dimensional_ME}) and numerical results in the full range of sizes studied. This is expected since, unlike the previous case, the analytical prediction also includes the higher order terms.

\section{Gap size dependence by periodic orbit theory}\label{sec3}
In this section we compute analytically the size dependence of the superconducting gap, including contributions from matrix elements and density of states, by using periodic orbit theory. The final expression for the gap captures quantitatively oscillations induced by shell effects. We refer to \cite{Brack1997} for a pedagogical introduction to this technique though we do provide a brief summary in appendix \ref{Ap_PO}. The square and cube cases are discussed separately.

\subsection{Three Dimensions}
The starting point is to re-write the gap equation by substituting  Eq. (\ref{matrixEls}) into Eq. (\ref{GapEqn}),
\begin{widetext}
\begin{equation}\label{GapEq2}
\Delta_n=\frac{\lambda}{2}\displaystyle\int_{-\epsilon_D}^{\epsilon_D}\frac{\Delta_{n'}\left(\nu_3(\epsilon')+\sum\limits_{r\in\{x,y,z\}}\frac{\nu_2 (\epsilon''_{r})}{2}+\sum\limits_{\langle r,s\rangle}\frac{\nu_1(\epsilon'''_{r,s})}{4}+\frac{\delta(\epsilon')}{8}\right)}{\sqrt{\epsilon'^2+\Delta_{n'}^2}\nu(0)}d\epsilon'
\end{equation}
\end{widetext}
where $\nu_i(\epsilon)$  is the density of states at energy $\epsilon$ in a cube-like billiard of size $L$ in $i$ dimensions and, 
\begin{equation}
\begin{cases}
\epsilon''_r=\epsilon'-\frac{\hbar^2\pi^2}{2mL^2}n_r^2\\
\epsilon'''_{r,s}=\epsilon'-\frac{\hbar^2\pi^2}{2mL^2}(n_r^2+n_s^2)
\end{cases}
\end{equation}
The density of states in a finite-size systems can be written as \cite{Brack1997},
\begin{equation}
\nu(\epsilon)=\nu_{TF}(\epsilon)(1+\bar{g}(\epsilon)+\tilde{g}_l(\epsilon))
\end{equation}
 where $\nu_{TF}(0)$ is the bulk Thomas-Fermi density of states, $\bar{g}(\epsilon)=-3\pi/2k_FL+ \ldots$ is the monotonous contribution, usually referred to as Weyl's expansion where we assume Dirichlet boundary conditions. Finally $\tilde{g}(\epsilon)$ is the oscillating contribution which can be expressed as a sum over periodic orbits of the classical counterpart. See appendix \ref{Ap_PO} for explicit expressions of $\nu_i$. 
Using the ansatz $\Delta=\Delta_0(1+f^{(1)}+f^{(3/2)}+f^{(2)}+ \ldots)$ 
with $f^{(k)}\propto (k_F L)^{-k} $ we expand the gap equation in powers of the small parameter $(k_FL)^{-1/2}$ and solve order by order to find,
\begin{equation}
\begin{split}
 f&^{1}=\frac{1}{2}\int^{\epsilon_D}_{-\epsilon_D}\frac{\Gamma^1 d\,\epsilon'}{\sqrt{\epsilon'^2+\Delta_0^2}}\\
 f&^{3/2}=\frac{1}{2}\int^{\epsilon_D}_{-\epsilon_D}\frac{\Gamma^{3/2} d\,\epsilon'}{\sqrt{\epsilon'^2+\Delta_0^2}}\\
 f&^{2}=\frac{1}{2}\int^{\epsilon_D}_{-\epsilon_D}\frac{\Gamma^2 d\,\epsilon'}{\sqrt{\epsilon'^2+\Delta_0^2}}+\frac{1}{2}(f^{1})^2\\&-\frac{\Delta_0^2f^1}{2}\int^{\epsilon_D}_{-\epsilon_D}\frac{\Gamma^1 d\,\epsilon'}{(\epsilon'^2+\Delta_0^2)^{3/2}}
\end{split}
\end{equation}
where we have collected terms in the numerator according to their $k_FL$ dependence such that $\Gamma^k\propto (k_F L)^{-k}$.
Applying the asymptotic form of the Bessel function $J_0(x)=\sqrt{\frac{2}{\pi x}}\cos(x-\frac{\pi}{4})$, expanding $\epsilon$ about the Fermi energy and carrying out the integrals, we arrive at the following expression for the gap, 
\begin{widetext}
\begin{align*}\label{gapwith3d}
 f^{(1)}=&\sum_{L_{\bf n}\neq 0}^\infty j_0(k_F L_{\bf{n}})\omega^{(1/2)}(L_{\bf n})\\
 f^{(3/2)}=&\frac{\pi}{2k_FL}\sum_{L_{\bf n}\neq 0}^\infty\left(\sum_{r\in\{x,y,z\}}J_0(X_rk_F L_{\bf{n}}^{i,j})\omega^{(1/2)}\left(\frac{L_{\bf n}^{i,j}}{X_{r}}\right)-3 J_0(k_F L_{\bf{n}}^{i,j})\omega^{(1/2)}(L_{\bf n}^{i,j})  \right)\\
 f^{(2)}=&\frac{\pi}{(k_FL)^2\lambda}\left(\sum_{\langle r,s\rangle}\frac{1}{2X_{r,s}}-\sum_{r\in\{x,y,z\}}\frac{1}{X_r}\right)+f^{(1)}\left(\frac{f^{(1)}}{2}-\sum_{L_{\bf n}\neq 0}^\infty j_0(k_F L_{\bf{n}})\omega^{(3/2)}(L_{\bf n})\right)  \numberthis 
\\ &+\frac{\pi}{(k_FL)^2}\sum_{L_{\bf n}\neq 0}^\infty \left( 3\cos(k_F L_{\bf{n}}^{i} )\omega^{(1/2)}(L_{\bf n}^{i}) -\sum_{r\in\{x,y,z\}}\frac{2}{X_r} \cos(X_rk_F L_{\bf{n}}^{i} )\omega^{(1/2)}\left(\frac{L_{\bf n}^{i}}{X_r}\right)\right. \\
&+\left.\sum_{\langle r,s\rangle}\frac{1}{2X_{r,s}} \cos(X_{r,s}k_F L_{\bf{n}}^{i} )\omega^{(1/2)} \left(\frac{L_{\bf n}^{i}}{X_{r,s}}\right)\right)
\end{align*}
where  $X_r=\sqrt{1-\left(\frac{\pi n_r}{k_FL}\right)^2}$, $X_{r,s}=\sqrt{1-\left(\frac{\pi}{k_FL}\right)^2(n_r^2+n_s^2)}$ and the weight functions $\omega$ are given by, 

\begin{equation}
\begin{split}
\omega^{(1/2)}(L_{\bf n})=\frac{1}{2}\int^\infty_{-\infty}\frac{\cos\left(\frac{L_{\bf n} t}{\zeta}\right)}{\sqrt{1+t^2}}d\,t=K_0\left(\frac{L_{\bf n}}{\zeta}\right)\\
\omega^{(3/2)}(L_{\bf n})=\frac{1}{2}\int^\infty_{-\infty}\frac{\cos\left(\frac{ L_{\bf n} t}{\zeta}\right)}{(1+t^2)^{3/2}}d\,t=\frac{L_{\bf n}}{\zeta}K_1\left(\frac{L_{\bf n}}{\zeta}\right)
\end{split}
\end{equation}
where $K_j$ is the modified Bessel function of the second kind of order $j$, $\zeta=\hbar^2k_F/m\Delta_0$ is the coherence length. These weight functions suppress exponentially the contribution of periodic orbits $L_{\bf n}$ longer than the coherence length $\zeta$.

For the sake of comparison we have also derived the analytical expression of the gap size dependence in the limit $I=1$ first obtained in Ref.\cite{Garcia-Garcia2011} (see Eq.($18$)-($19$)),
\begin{equation}\label{right2011}
\begin{split}
 &f_{I=1}^{(1)}=-\frac{3\pi}{2\lambda}\frac{1}{k_FL}+\sum_{L_{\bf n}\neq 0}^\infty j_0(k_F L_{\bf{n}})\omega^{(1/2)}_1(L_{\bf n})\\
 &f_{I=1}^{(3/2)}=-\frac{3\pi}{2k_FL}\sum_{L_{\bf n}\neq 0}^\infty J_0(k_F L_{\bf{n}}^{i,j})\omega^{(1/2)}_1(L_{\bf n}^{i,j})  \\
 &f_{I=1}^{(2)}=f_{I=1}^{(1)}\left(\frac{f_{I=1}^{(1)}}{2}+\frac{3\pi}{2}\frac{1}{k_FL}-\sum_{L_{\bf n}\neq 0}^\infty j_0(k_F L_{\bf{n}})\omega^{(3/2)}(L_{\bf n})\right)+\frac{3\pi}{(k_FL)^2}\sum_{L_{\bf n}\neq 0}^\infty  \cos(k_F L_{\bf{n}}^{i} )\omega^{(1/2)}_1(L_{\bf n}^{i}). \\
\end{split}
\end{equation}
\end{widetext}
 
The expansion above does not agree completely with that of Ref.\cite{Garcia-Garcia2011}. There is a factor of $\frac{1}{2}$ missing before the $f^{(1)2}$ term on the first line of Eq. ($19$) of Ref.\cite{Garcia-Garcia2011}. On the second line, the term $\tilde{g}^{(1)}$ should be replaced by $\tilde{g}^{(3)}$. Finally in the equation for $W_{3/2}(L_P/\xi)$, just below Eq. $(20)$ of Ref.\cite{Garcia-Garcia2011}, the pre-factor $\Delta_0^2$ should be replaced by $\lambda$. 

 It is also important to note a crucial limitation of the semi-classical expansion in the small variable $(k_FL)^{-1}\ll 1$, not discussed in Ref.\cite{Garcia-Garcia2011}, which is especially relevant in the case of symmetric grains. From Eq.(\ref{right2011}) it is clear that the pre-factors in front of the expansion parameter involve sums over all periodic orbits shorter than the coherence length. For typical values of parameters $\xi \sim 200$nm and $L \sim 10$nm the sum runs over thousands periodic orbits. It is entirely plausible that for sizes for which shell effects are strong, and therefore the different terms of the oscillating sum add coherently, these pre-factors can become very large to the point that $f^{(i)}>1$ and the semiclassical expansion breaks down. The exact range of validity of the expansion is going to be very sensitive to the choice of parameters since $\xi \propto e^{1/\lambda}$. For instance we have found that for $\lambda > 0.3$ and $E_D \sim 30$meV  it will be convergent for almost all sizes $L > 15$nm. For $\lambda > 0.4$, and the same $E_D$, it will converge for all sizes $L > 10$nm.

Here our main goal is to study analytically the role of the matrix elements in the semiclassical expansion. Therefore we use the two expressions above to find the difference between the size dependence of the superconducting gap with Eq.(\ref{gapwith3d}) and without Eq.(\ref{right2011}) non-trivial matrix elements,
\begin{widetext}
\begin{equation}\label{3dgapdif}
\begin{split}
&\Delta_{\text{Diff}}=\frac{\Delta_{\text{Exact}}-\Delta_{I=1}}{\Delta_0}= f_\text{Diff}^{(1)}+f_\text{Diff}^{(3/2)}+f_\text{Diff}^{(2)} \\
& f_\text{Diff}^{(1)}=\frac{3\pi}{2\lambda}\frac{1}{k_FL}\\
& f_\text{Diff}^{(3/2)}=\frac{\pi}{2k_FL}\sum_{L_{\bf n}\neq 0}^\infty\sum_{r\in\{x,y,z\}}J_0(X_rk_F L_{\bf{n}}^{i,j})\omega^{(1/2)}\left(\frac{L_{\bf n}^{i,j}}{X_{r}}\right)\\
& f_\text{Diff}^{(2)}=\frac{\pi}{(k_FL)^2\lambda}\left(\sum_{\langle r,s\rangle}\frac{1}{2X_{r,s}}-\sum_{r\in\{x,y,z\}}\frac{1}{X_r}\right)+\left(\frac{3\pi}{2}\frac{1}{k_FL}\right)^2\left(\frac{1}{\lambda}-\frac{1}{2\lambda^2}\right)\\
&+\frac{3\pi}{2k_FL}\sum_{L_{\bf n}\neq 0}^\infty\left(\left(\frac{1}{\lambda}-1\right)j_0(k_F L_{\bf{n}})\omega^{(1/2)}(L_{\bf n})-\frac{1}{\lambda} j_0(k_F L_{\bf{n}})\omega^{(3/2)}(L_{\bf n})\right)\\
&+\frac{\pi}{(k_FL)^2}\sum_{L_{\bf n}\neq 0}^\infty \left(\sum_{\langle r,s\rangle}\frac{1}{2X_{r,s}} \cos(X_{r,s}k_F L_{\bf{n}}^{i} )\omega^{(1/2)} \left(\frac{L_{\bf n}^{i}}{X_{r,s}}\right)-\sum_{r\in\{x,y,z\}}\frac{2}{X_r} \cos(X_rk_F L_{\bf{n}}^{i} )\omega^{(1/2)}\left(\frac{L_{\bf n}^{i}}{X_r}\right)\right).
\end{split}
\end{equation}
\end{widetext}

\begin{figure}[H]
    \includegraphics[width=0.42\textwidth]{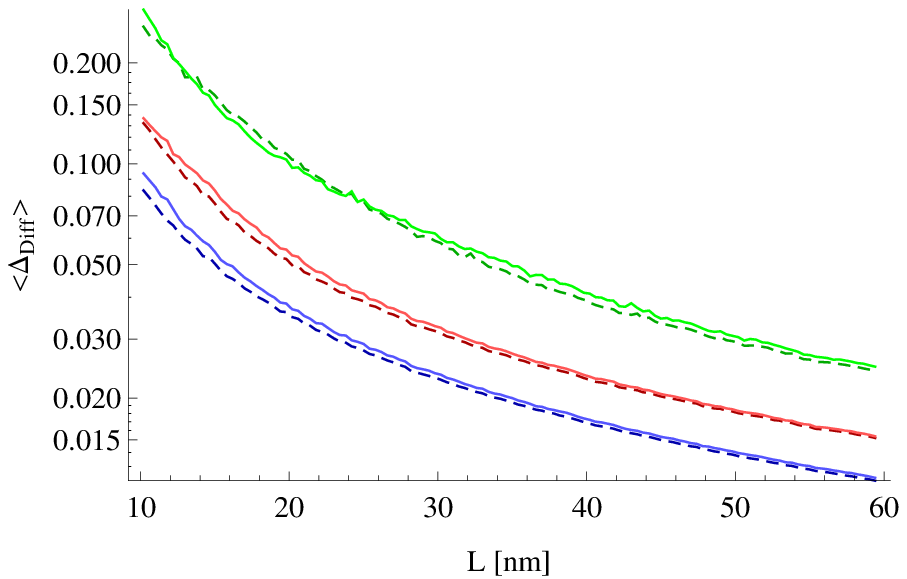}
    \includegraphics[width=0.42\textwidth]{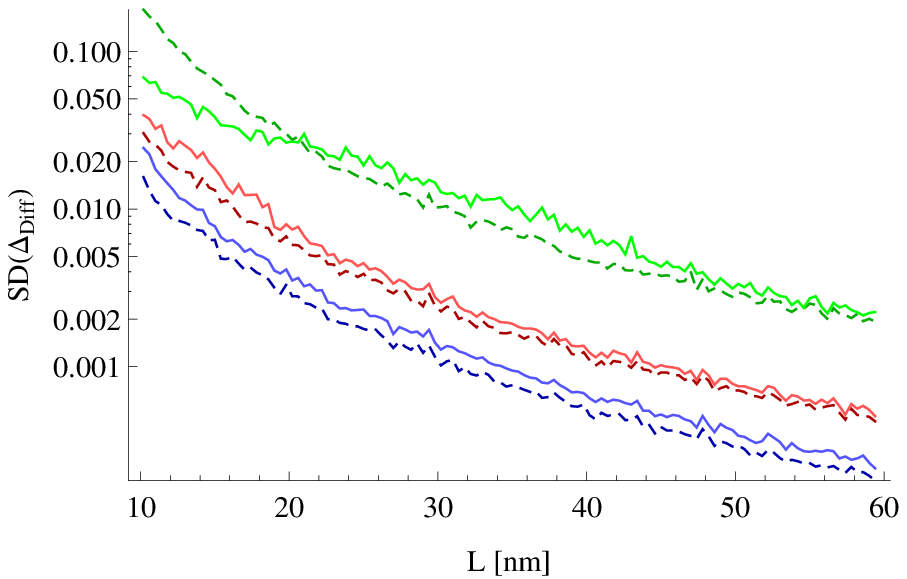}
 \caption{Comparison of the numerical and analytical calculation of $\Delta_{\text{Diff}}$, the difference between the superconducting gap with and without matrix elements. The upper plot shows the mean value and the lower plot the standard deviation taken over consecutive intervals of size $0.4$nm. The solid line shows numerical results and the dashed line the results from the periodic orbit calculation, Eq.(\ref{3dgapdif}). From top to bottom the line pairs correspond to $\lambda=0.2$ (green),$0.3$ (red),$0.4$ (blue). We note the extremely good agreement not just in the line shape but also in the fine structure of the standard deviation. The contribution of the matrix elements to the gap size dependence is substantial in the region $L \sim 10$nm and where the BCS formalism is still applicable. Deviations are still noticeable even for much larger grains $L \leq 50$nm. As was expected (see text) the expansion begins to breakdown for the case of $\lambda=0.2$, $L\sim 10$nm.} \label{PO3d}
\end{figure} 
The limitation of the semiclassical expansion due to shell effects mentioned above also applies to Eq.(\ref{3dgapdif}) but there are important differences. The potentially most divergent sums cancel each other which increases substantially the convergence of the expansion. For $\lambda=0.3$ and $E_D=30$meV the expansion is convergent for almost all $L > 10$. Indeed, as can be observed in figure \ref{PO3d}, the agreement between the numerical and analytical results is excellent for any $\lambda > 0.3$ and $L > 10$nm. We also note that for $L \sim 10$nm, $\Delta_{\text{Diff}}$, that physically describes the contribution of the matrix elements to the superconducting gap, is substantial. That suggests that any quantitative description of superconductivity in nano-grains must take it into account.

 We note that deviations for smaller $\lambda$ is an indication of the incipient breaking of the semiclassical expansion due to strong shell effects. Indeed we have checked that, in this case, including higher orders in the expansion only worsens the agreement with the numerical results. In part this is also due to the presence of crossed terms $f^{(1)}f^{(3/2)}$ 
 which, despite being of higher order in $(k_FL)^{-1}$, have the potential to be larger than those of lower order at sizes for which shell effects are important. 

A natural question to ask is whether these results are really relevant for realistic superconducting grains. It is reasonable to neglect disorder since current growth techniques make it feasible to reach mean free paths much larger than the grain size. Small deviations from a highly symmetric geometry, due to imperfections, can be included in the semiclassical formalism by adding an additional cutoff length that describes the typical length that a particle travels inside the grain without hitting the imperfection. If this length is larger than the coherence length it has no impact at all on our results. If it is shorter it will diminish shell effects but their role will still be important provided the imperfection scattering length is much larger than the grain size. Highly symmetric grains such as hemispheres \cite{Bose2010} are within the reach of current experimental techniques however we are not yet aware of experimental results regarding nano-cubes or nano-squares.

\subsection{Two Dimensions}
Following the same prescription as in the three dimensional case we use the ansatz $\Delta=\Delta_0(1+f^{1/2}+f^{1}+\ldots)$ and solve order by order. In the absence of matrix elements we find,
\begin{widetext}
\begin{equation}
\begin{split}
&f^{1/2}_{I=1}=\sum_{L_{\bf n}}J_0(k_FL^{ij}_{\bf n})\omega^{(1/2)}(L^{ij}_{\bf n})\\
&f^{1}_{I=1}=f^{(1/2)}_{I=1}\left(\frac{f^{(1/2)}_{I=1}}{2}-\sum_{L_{\bf n}}J_0(k_FL^{ij}_{\bf n})\omega^{(3/2)}(L^{ij}_{\bf n})\right)-\frac{2}{\lambda k_FL}-\frac{4}{k_FL}\sum_{L_{\bf n}}\cos(k_FL^i_{\bf n})\omega^{(1/2)}(L^i_{\bf n}).
\end{split}
\end{equation}
With the matrix elements included we find,
\begin{equation}
\begin{split}
f^{1/2}=&\sum_{L_{\bf n}}J_0(k_FL^{ij}_{\bf n})\omega^{(1/2)}(L^{ij}_{\bf n})\\
f^{1}=&f^{(1/2)}\left(\frac{f^{(1/2)}}{2}-\sum_{L_{\bf n}}J_0(k_FL^{ij}_{\bf n})\omega^{(3/2)}(L^{ij}_{\bf n})\right)+\frac{1}{\lambda k_FL}\left(\sum_{r\in\{x,y\}}\frac{1}{X_r}-2\right)-\frac{4}{k_FL}\sum_{L_{\bf n}}\cos(k_FL^i_{\bf n})\omega^{(1/2)}(L^i_{\bf n})\\
&+\sum_{r\in\{x,y\}}\frac{1}{X_rk_FL}\sum_{L_{\bf n}}\cos(X_rk_FL^i_{\bf n})\omega^{(1/2)}(L^i_{\bf n}/X_r).
\end{split}
\end{equation}
The difference between the two leads to the final expression for the gap size corrections due to matrix elements,
\begin{equation}
\begin{split}\label{dif2d}
\Delta_\text{diff}=\frac{1}{\lambda k_FL}\sum_{r\in\{x,y\}}\frac{1}{X_r}+\sum_{r\in\{x,y\}}\frac{1}{X_rk_FL}\sum_{L_{\bf n}}\cos(X_rk_FL^i_{\bf n})\omega^{(1/2)}(L^i_{\bf n}/X_r).
\end{split}
\end{equation}
\end{widetext}

In this case we have also found a very good agreement between Eq.(\ref{dif2d}) and numerical results.

Naively one might expect the final number theory results Eq. (\ref{2d_num_res}) and the periodic orbit results Eq. (\ref{dif2d}) to be similar. On first inspection though, they appear to be quite different, in particular it is not clear where the logarithm Eq. (\ref{2d_num_res}) can be found in the semi-classical expressions. The relationship between the number theory and periodic orbit results is not entirely straightforward however as in the former we are studying the smoothed value of the matrix element whereas for periodic orbits we have calculated the difference in the superconducting gap with and without matrix element. In principle it should be possible to derive the number theory results from the semi-classical density of states by taking care to include the smoothing over the Debye window for $X_r, X_{r,s}$ terms. This task would be difficult however without applying results from number theory.

\section{Conclusions}
We have computed analytically the size dependence of the energy gap for square and cubic superconducting grains in the mean-field approximation by making extensive use of semi-classical and number theory techniques. For typical values of the parameters $\lambda \sim 0.3$, $E_D \sim 30$meV the result for the difference between the gap with and without matrix elements, our main finding, is in excellent agreement with numerical results for almost all sizes $L > 10$nm.  These results indicate that the contribution of the matrix elements to the superconducting gap is important to model superconductivity in the region $L \sim 10$nm. We note that mean-field-approaches are still valid in this region. For the square nano-grain, the expression for the average matrix elements has an inherent number theoretic nature. For the superconducting nano-cube the leading size correction is equal to the one for a chaotic grain which suggests that it is universal. 

\begin{acknowledgments}
JM acknowledges support from an EPSRC Ph.D. studentship. AMG was supported by EPSRC, grant No. EP/I004637/1, FCT, grant PTDC/FIS/111348/2009 and a Marie Curie International Reintegration Grant PIRG07-GA-2010-268172.
\end{acknowledgments}
\vspace{10mm}
\appendix

\section{Semiclassical Results}\label{Ap_PO}
Here we summarize the relevant Semiclassical results for the density of states\cite{Brack1997}. Applying the Gutzwillar trace formalism we may express the densities of states in the following form,
\begin{equation}
\nu(\epsilon)=\nu_{TF}(\epsilon)(1+\bar{g}(\epsilon)+\tilde{g}_l(\epsilon))
\end{equation}
Where $\nu_{TF}(0)$ is the Thomas-Fermi density of states in the bulk,
\begin{equation}
\nu_{TF}(\epsilon)=2\times
\begin{cases}
\frac{V}{4\pi^2}(\frac{2m}{\hbar^2})^{3/2}\sqrt{\epsilon+\epsilon_F}, &\text{(3D)}\\
\frac{A}{4\pi}(\frac{2m}{\hbar^2}), &\text{(2D)}\\
\frac{L}{2\pi}\sqrt{\frac{2m}{\hbar^2}}\frac{1}{\sqrt{\epsilon+\epsilon_F}}, &\text{(1D)}\\
\end{cases}
\end{equation}
$\bar{g}(\epsilon)$ is the smooth contribution, given by the Weyl expansion, in this work we have used Dirichlet boundary conditions,
\begin{equation}
\bar{g}(\epsilon)=
\begin{cases}
-\frac{\mathcal{S}\pi}{4kV}+\frac{2\mathcal{C}}{k^2 V}, &\text{(3D)}\\
-\frac{2L}{kA}, &\text{(2D)}\\
0, &\text{(1D)}\\
\end{cases}
\end{equation}
$\mathcal{S}$ is the surface area of the grain and $\mathcal{C}$ is the curvature. $\tilde{g}(\epsilon)$ is the oscillating contribution given by
\begin{equation}
\tilde{g}(\epsilon)=
\begin{cases}
\tilde{g}^{(3)}(\epsilon)-\frac{1}{2}\sum_i\sum_{j\neq i}\tilde{g}^{(2)}_{i,j}(\epsilon)+\frac{1}{4}\sum_i g_i^{(1)}(\epsilon), &\text{(3D)}\\
\tilde{g}^{(2)}_{1,2}(\epsilon)-\frac{1}{2}\sum_i g_i^{(1)}(\epsilon), &\text{(2D)}\\
g_1^{(1)}(\epsilon), &\text{(1D)}\\
\end{cases}
\end{equation}
These terms each correspond to the sum over a set of periodic orbits. $\tilde{g}^{(3)}(\epsilon)$ is over the orbits of length $L_{\bf n} =2\sqrt{L_x^2n_x^2+L_y^2n_y^2+L_z^2n_z^2}$ with $n_x,n_y,n_z$ not simultaneously zero, according to,
\begin{equation}
\tilde{g}^{(3)}(\epsilon)=\sum_{L_{\bf n}\neq 0}^\infty j_0(k L_{\bf{n}})
\end{equation}
where $j_0$ is the zeroth order spherical Bessel function. Similarly $\tilde{g}^{(2)}_{i,j}$ is over periodic orbits $L_{\bf n}^{i,j} =2\sqrt{L_i^2n_i^2+L_j^2n_j^2}$ 
\begin{equation}
\tilde{g}^{(2)}_{i,j}=
\begin{cases}
\frac{L_iL_j\pi}{k_FV}\sum_{L_{\bf n}\neq 0}^\infty J_0(k L_{\bf{n}}^{i,j}  ), &\text{(3D)}\\
\frac{L_iL_j}{A}\sum_{L_{\bf n}\neq 0}^\infty J_0(k L_{\bf{n}}^{i,j}  ), &\text{(2D)}\\
\end{cases}
\end{equation}
$J_0$ is the zeroth order Bessel function. $g_i^{(1)}$ sums over periodic orbits of length  $L_{\bf n}^{i} =2L_in_i$ 
\begin{equation}
\tilde{g}^{(1)}_{i}=
\begin{cases}
\frac{4\pi L_i}{k_F^2V}\sum_{L_{\bf n}\neq 0}^\infty \cos(k L_{\bf{n}}^{i} ), &\text{(3D)}\\
\frac{4L_i}{k_F A}\sum_{L_{\bf n}\neq 0}^\infty \cos(k L_{\bf{n}}^{i}  ), &\text{(2D)}\\
\sum_{L_{\bf n}\neq 0}^\infty \cos(k L_{\bf{n}}^{i} ), &\text{(1D)}\\
\end{cases}
\end{equation}
\begin{widetext}
\section{Number theory results}\label{ap:num}
{\it One Dimensional}\\
$r_1(n)$ is trivially,
\begin{equation}
r_1(n)=
\begin{cases}
1 &\text{if n is square}\\
0&\text{otherwise}
\end{cases}
\end{equation}

{\it Two Dimensional}\\
It has been shown that the number of representations of n as the sum of two squares is\cite{Gauss1801},
\begin{equation}\label{2sqTh}
r_2(n)=d_1(n)-d_3(n)
\end{equation}
where $d_l(n)$ is the number of divisors of $n$ congruent to $l (\bmod 4)$\\

{\it Three Dimensional}\\
It has been shown \cite{Bateman1951}, the number of representations of n as the sum of three squares is,
\begin{equation}
r_3(n)=\frac{\pi}{4} n^\frac{1}{2}\xi(3,n)\label{full3}
\end{equation}
where, 
\begin{equation}\nonumber
\xi(3,n)=\prod_p(1+A_p+A_{p^2}+\ldots)
\end{equation}
Where the product is over the prime divisors of n and these series truncate according to,
\begin{enumerate}[ label=(\alph{*})]
\item if $p=2, A_{2^a}$ is
\begin{enumerate}[ label=(\roman{*})]
\item $A_2=0$
\item if $a$ is even
\begin{equation}\nonumber
A_{2^a}=
\begin{cases}
\frac{\cos((\pi/4)(2n_1-3))}{2^{(a-1)/2}}&\text{if} \;\exists\; n_1\; \text{such that}\;  n=2^{a-2}n_1\;\;\;(n_1\; \text{not necessarily odd})\\
0 &\text{if}\; 2^{a-2}\nmid n
\end{cases}
\end{equation}
\item if $a$ is odd
\begin{equation}\nonumber
A_{2^a}=
\begin{cases}
(-1)^{(n_2-3)/4}2^{(a-1)/2}&\text{if} \;\exists\; n_2\; \text{such that}\;  n=2^{a-3}n_2\;\;\text{and}\; n\equiv 3 (\bmod 4)\\
0 & \text{otherwise}
\end{cases}
\end{equation}
\end{enumerate}
\item if $p$ is an odd prime,
\begin{enumerate}[ label=(\roman{*})]
\item if $a$ is even.
\begin{equation}\nonumber
A_{p^a}=
\begin{cases}
(p-1)p^{-(a/2+1)}i^{ 3\left(\frac{p^a-1}{2}\right)^2}& \text{if}\;p^a\mid n\\
-p^{-(a/2+1)}i^{3\left(\frac{p^a-1}{2}\right)^2}&\text{if}\;p^{a-1}\mid\mid n\\
0 &\text{if}\;p^{a-1}\nmid n
\end{cases}
\end{equation}
\item $a$ is odd,
\begin{equation}\nonumber
A_{p^a}=
\begin{cases}
0 & \text{if either} \;  p^a\mid n \text{ or } p^{a-1}\nmid n \\
p^{-(a+1)/2}\left(\frac{n_1}{p}\right)_J i^{3\left(\frac{p^a-1}{2}\right)^2}\frac{(1-i)(1+i^p)}{2}&\text{such that } n=p^{a-1}n_1\text{ and }p\nmid n_1
\end{cases}
\end{equation}
\end{enumerate}
\end{enumerate}
Where $\left(\frac{a}{b}\right)_J$ is the Jacobi Symbol. So, for example, in the case where $n$ is odd and square free\cite{Grosswald1985},
\begin{equation}
r_3(n)=\xi\frac{2\sqrt{n}}{\pi}\sum_{m=1,m \text{ odd}}^\infty\left(\frac{-n}{m}\right)_J\frac{1}{m}
\end{equation}
\begin{equation}\nonumber
\xi=1+\frac{1}{\sqrt{2}}\cos(\pi(2n-3)/4)+\frac{1}{2}\cos(\pi(n-3)/4)
\end{equation}
\end{widetext}
\vspace{5mm}
\bibliography{library,footnore}
\end{document}